\documentclass[aps,pre,reprint,superscriptaddress,longbibliography]{revtex4-1}
\usepackage{amsmath,amssymb,graphicx,subfigure,color,times,tabularx}

\begin{document}
\title{\bf {Non-unique Hamiltonians for Discrete Symplectic Dynamics}}
\author{Liyan Ni}
\affiliation{Institute of Frontier Chemistry, School of Chemistry and Chemical Engineering, Shandong University, Qingdao,   266237,  P. R. China}
\affiliation{Qingdao Institute for Theoretical and Computational Sciences (QiTCS), Shandong University, Qingdao, 266237, P. R. China}
\author{Yihao Zhao}
\affiliation{Institute of Frontier Chemistry, School of Chemistry and Chemical Engineering, Shandong University, Qingdao,   266237,  P. R. China}
\affiliation{Qingdao Institute for Theoretical and Computational Sciences (QiTCS), Shandong University, Qingdao, 266237, P. R. China}
\author{Zhonghan Hu} \email{zhonghanhu@sdu.edu.cn}
\affiliation{Institute of Frontier Chemistry, School of Chemistry and Chemical Engineering, Shandong University, Qingdao,   266237,  P. R. China}
\affiliation{Qingdao Institute for Theoretical and Computational Sciences (QiTCS), Shandong University, Qingdao, 266237, P. R. China}

\begin{abstract}
An outstanding property of any Hamiltonian system is the symplecticity of its flow, namely, the continuous trajectory preserves volume in phase space.
Given a symplectic but discrete trajectory generated by a transition matrix applied at a fixed time-increment ($\tau > 0$), it was generally believed that there exists a unique Hamiltonian producing a continuous trajectory that coincides at all discrete times ($t = n\tau$ with $n$ integers) as long as $\tau$ is small enough.
However, it is now exactly demonstrated that, for any given discrete symplectic dynamics of a harmonic oscillator, there exist an infinite number of real-valued Hamiltonians for any small value of $\tau$ and
an infinite number of complex-valued Hamiltonians for any large value of $\tau$. 
In addition, when the transition matrix is similar to a Jordan normal form with the supradiagonal element of $1$ and the two identical diagonal elements of either $1$ or $-1$, 
only one solution to the Hamiltonian is found for the case with the diagonal elements of $1$, but no solution can be found for the other case.
\end{abstract}
\maketitle

Symplectic integrators\cite{DeVogelaere1956,Ruth1983,Feng1985,Feng_Qin1987} are widely used to simulate the dynamic processes of elementary particles, materials and celestial bodies\cite{Yoshida1990,Frenkel_Smit2023}.
In order to understand the structure, regularity and stability of the discrete dynamics generated by these integrators, it may be worth analyzing the motion in a Hamiltonian representation, just as it is realized in classical mechanics\cite{Arnold1989}.
Despite recent progress\cite{Griffiths_Sanz-Serna1986,Friedman1991,Yoshida1993,Toxvaerd1994,Hairer1994,Chin_Scuro2005,Toxvaerd_Dyre2012,Hammonds_Heyes2020}, the basic problems about the uniqueness and existence of this representation have not been solved for any model system.

When the classical system described by an original Hamiltonian, ${\cal H}_0(q,p)$, is propagated at a fixed time-increment $\tau$:
\begin{equation}  \begin{bmatrix} q((n+1)\tau) \\   p((n+1)\tau) \end{bmatrix} = \hat{R}  \begin{bmatrix}  q(n\tau) \\ p(n\tau) \end{bmatrix},  \label{eq:dd} \end{equation}
with $n=0,1,2,\cdots$ an integer and $\hat{R}$ a symplectic transition matrix derived from ${\cal H}_0(q,p)$,
it is generally believed that there exists a slightly perturbed Hamiltonian ${\cal H}(\tau,q,p)$, explained clearly in the review by Yoshida\cite{Yoshida1993}, 
such that the discrete phase points, $(q(n\tau),p(n\tau))$, lie on the continuous trajectory produced by Hamilton's canonical equations of motion,
\begin{equation} \left\{ \begin{aligned} \frac{d q(t)}{dt} \equiv \dot{q} &= \dfrac{\partial {\cal H}}{\partial p}  \\  \frac{d p(t)}{dt} \equiv \dot{p} &= -  \dfrac{\partial {\cal H}}{\partial q} \end{aligned} \right.\label{eq:ceh0}. \end{equation}
${\cal H}(\tau,q,p)$ was previously expressed by a formal power series in $\tau$ (e.g. Eq. (45) of ref.\cite{Yoshida1993}):
\begin{equation} {\cal H}(\tau,q,p) = {\cal H}_0(q,p) + \tau {\cal H}_1(q,p) + \tau^2 {\cal H}_2(q,p) + \cdots. \label{eq:hseries}\end{equation}
When $\tau$ approaches $0$, ${\cal H}(\tau,q,p)$ necessarily reduces to ${\cal H}_0(q,p)$ up to a trivial additive constant independent of $q$ and $p$.
Higher order corrections are uniquely formulated in terms of the Baker-Campbell-Hausdorff (BCH) expansion\cite{Varadarajan1984} for the product of exponential operators involved in $\hat{R}$\cite{Dragt_Finn1976,Dragt1988,Yoshida1993}.
In this work, we instead solve exactly the system of one harmonic oscillator to explicitly demonstrate that, contrary to the assumed uniqueness, ${\cal H}(\tau,q,p)$ is in fact non-unique even for small $\tau$.
\begin{figure}[!htb]\centerline{\includegraphics[width=8cm]{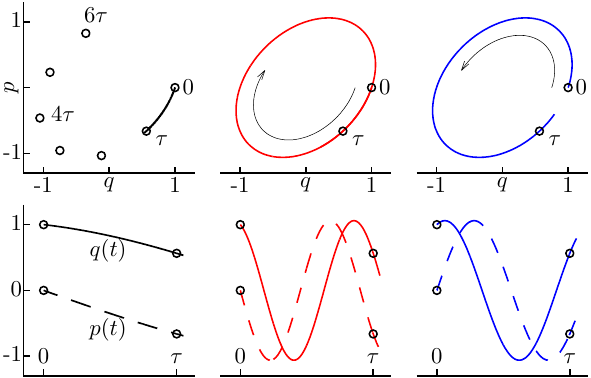} }
\caption{Typical continuous trajectories from $(q(0)=1, p(0)=0)$ to $(q(1.05\tau),p(1.05\tau))$ in the $qp$ phase space (top) and as functions of time (bottom, solid lines for $q(t)$ and dashed $p(t)$) generated by the Hamiltonians of Eq.~\eqref{eq:h1} at $m=0$ (left), $1$ (middle, red) and $-1$ (right, blue).
In the phase space, the trajectory of the phase points rotates clockwise ($m \geqslant 0$) and counter-clockwise ($ m < 0$) respectively.
Discrete points: $q(n\tau)$ and $p(n\tau)$ at $n=0,1,2,\cdots, 6$ (circles), are generated by the symplectic integrator. See Eq.~\eqref{eq:ct}.}
\label{fig:1}\end{figure}

For the system of the one-dimensional (1D) single harmonic oscillator defined by ${\cal H}_0=q^2/2 + p^2/2$ in reduced units, the non-singular $2$ by $2$ transition matrix: 
\begin{equation} \hat{R}(\tau) =  \begin{bmatrix}   R_1(\tau) & R_2(\tau)  \\ R_3(\tau)  & R_4(\tau) \end{bmatrix} , \end{equation}
depends on none of $n$, $q$ and $p$; it is only a function of the time-increment $\tau$.
The symplectic condition requires that the determinant of $\hat{R}$ is unity: $R_1R_4 - R_2R_3 = 1$, so the product of its two eigenvalues is $1$. 
Therefore, we denote the eigenvalues as $y$ and $1/y$, where $y$ is a complex number expressed in the exponential form: $y = \left|y\right| e^{i\theta}$ with $-\pi < \theta \leqslant \pi$.
Possible solutions to ${\cal H}(\tau,q,p) $, which are independent of time or $n$, are listed in the following three categories.

i) When $(R_1+R_4)^2 \neq 4 $, $\hat{R}$ has two distinct eigenvalues: $y\neq 1/y$, and ${\cal H}$ takes a binomial form:
\begin{equation} {\cal H}(\tau, q, p|m) = \frac{\log(y,m)}{\left(y - 1/y \right) } \frac{ R_2 p^2 - R_3  q^2 + (R_1-R_4) pq }{\tau}, \label{eq:h1} \end{equation}
where the multivalued logarithm function is defined as
\begin{equation} \log(y,m) = \log \left| y \right| + i \theta + i 2m\pi ,  \label{eq:log1} \end{equation}
with $m$ an arbitrary integer and $i$ the imaginary unit.
i-a) When $\theta\neq 0$ and $\theta\neq \pi$, which is often the case for small $\tau$ in the traditional symplectic integrators\cite{Yoshida1990,Frenkel_Smit2023},
the two distinct eigenvalues are complex conjugated: $ 1/y = y^*$ and the module must be $1$: $\left| y \right|^2 = y y^* = 1$.
Both $\log(y,m) $ and $y-1/y$ give pure imaginary numbers and consequently ${\cal H}(\tau, q, p | m )$ are all real-valued. 
Specifically, ${\cal H}(\tau, q, p | 0 )$ reduces to the power series of Eq.~\eqref{eq:hseries} applied to the harmonic oscillator.
Typical trajectories in the phase space and as functions of time for $m=0,\pm 1$ are shown in Fig.~\ref{fig:1}.
i-b) When $\theta = 0$, the two distinct eigenvalues are both positive: $y>0$, and then ${\cal H}(\tau, q, p |0 )$ is still real-valued and other solutions with $m\neq 0$ all complex.
i-c) When $\theta=\pi$, which is often the case for large $\tau$, the two distinct eigenvalues are both negative: $y<0$, and then no real-valued ${\cal H}(\tau, q, p |m )$ exists.

ii) When $\hat{R} = \pm \hat{I}$ with $\hat{I}$ the $2$ by $2$ identity matrix, $y = 1/y = \pm 1$ and then ${\cal H}$ exists:
\begin{equation} {\cal H}(\tau, q, p|m) = i\pi \frac{ 4m + 1 \mp 1 }{2} \frac{  C_2 p^2 - C_3  q^2 + 2C_1 pq }{2\tau}, \label{eq:h2} \end{equation}
with $C_1$, $C_2$ and $C_3$ arbitrary complex numbers satisfying $C_1^2 + C_2C_3 = 1$. There are an infinite number of real-valued and complex-valued Hamiltonians.

iii) When $y = 1/y = \pm 1$ and $\hat{R}$ is similar to a Jordan normal form with the supradiagonal element of $1$ and the diagonal elements of $\pm 1$:
\begin{equation} \hat{R} = \hat{P} \begin{bmatrix}   \pm 1 & 1  \\ 0  & \pm 1 \end{bmatrix} \hat{P}^{-1}, \label{eq:jordan} \end{equation}
with $\hat{P}$ non-singular and $\hat{P}\hat{P}^{-1} = \hat{I}$, there must be $R_1+R_4 = \pm 2$ but $\hat{R}\neq \pm\hat{I}$.  A unique or no solution to ${\cal H}$ is found respectively. iii-a) When $y=1/y=1$ and $\hat{R}\neq \hat{I}$, the unique solution is 
\begin{equation} {\cal H}(\tau, q, p) = \frac{ R_2 p^2 - R_3 q^2  + 2 (R_1 - 1) pq }{2\tau}. \label{eq:h3} \end{equation}
iii-b)  When $y = 1/y = -1$ and $\hat{R} \neq -\hat{I}$, no solution to ${\cal H}$ can be found.

{\it Derivation.} Obviously, the phase point that coincides with Eq.~\eqref{eq:dd} at any discrete time could evolve continuously according to
\begin{equation}  \begin{bmatrix} q(t) \\   p(t) \end{bmatrix} = {\hat{R}}^{t/\tau} \begin{bmatrix}  q(0) \\ p(0) \end{bmatrix}= \exp\left(\dfrac{t}{\tau}\hat{Z}\right)  \begin{bmatrix}  q(0) \\ p(0) \end{bmatrix} , \label{eq:expm} \end{equation}
where the matrix $\hat{Z}$ is determined by the matrix equation,
\begin{equation} e^{\hat{Z}} = \hat{R},  \label{eq:logR}\end{equation}
with the exponential interpreted as a Taylor series:
\begin{equation} e^{\hat{Z}} = \hat{I}  + \hat{Z} + \frac{1}{2}\hat{Z}^2 + \cdots + \frac{1}{k!}\hat{Z}^k + \cdots . \label{eq:ez} \end{equation}
For any known $\hat{R}$, all the solutions of Eq.~\eqref{eq:logR} are called (natural) logarithm of $\hat{R}$ (p. 239 of ref.\cite{Gantmacher1959}). 
The eigenvalues $x_j$ of $\hat{Z}$ are connected with the eigenvalues $y_j$ of $\hat{R}$ by the formula: $y_j = e^{x_j}$, thus, $y_j$ must be non-zero, i.e., $\hat{R}$ is non-singular, such that $x_j$ exists. 
In addition, the Hamilton-Cayley theorem states that any matrix satisfies its own characteristic equation (p.83 of ref.\cite{Gantmacher1959}):
\begin{equation} \left( Z_1 \hat{I} - \hat{Z}\right)\left( Z_4 \hat{I} - \hat{Z} \right)  - Z_2Z_3 \hat{I} = 0. \end{equation}
This theorem simplifies all higher order multiplications in the Taylor series of Eq.~\eqref{eq:ez} to linear combinations of $\hat{Z}$ and $\hat{I}$ only. 
As a consequence, Eq.~\eqref{eq:logR} implies a linear matrix equation with two coefficients, $a$ and $b$:
\begin{equation} a\hat{Z} + b\hat{I} = \hat{R} . \label{eq:coeff}\end{equation}

The connection between the assumed continuous trajectory of Eq.~\eqref{eq:expm} and the canonical equations of motion, Eq.~\eqref{eq:ceh0}, is made clear by taking the time derivative:
\begin{equation} \left\{ \begin{aligned} \dot{q}  = \frac{Z_1 q + Z_2 p}{\tau} &= \dfrac{\partial {\cal H}}{\partial p}  \\  \dot{p}  = \frac{Z_3 q + Z_4 p}{\tau}  &= -  \dfrac{\partial {\cal H}}{\partial q} \end{aligned} \right.\label{eq:ceh2}. \end{equation}
Certainly, the elements of $\hat{Z}$ for the harmonic oscillator are only functions of $\tau$ for the reason that $\hat{R}$ is. 
For the partial derivatives being linearly dependent on $p$ and $q$, ${\cal H}$ must exist in a binomial form up to an additive constant:
\begin{equation} {\cal H} = \frac{ Z_2 p^2 - Z_3 q^2 + 2 Z_1 pq }{2 \tau}  =  \frac{ Z_2 p^2 - Z_3 q^2 - 2 Z_4 pq }{2 \tau},  \label{eq:hgeneral} \end{equation}
if and only if $\hat{Z}$ satisfying Eq.~\eqref{eq:logR} is traceless: $Z_1 + Z_4 = 0$, or equivalently, the sum of the two eigenvalues of $\hat{Z}$ equals $0$: $x_1 + x_2 = 0$. 
The symplectic condition imposed on $\hat{R}$: $ R_1R_4 - R_2R_3 = y_1y_2 = e^{x_1 + x_2} = 1$ becomes necessary, as expected.
However, as demonstrated in the below, this condition is not always sufficient to reach a traceless $\hat{Z}$ from Eq.~\eqref{eq:logR}.

By analyzing the elementary divisors (Jordan blocks) of $\hat{R}$ and then expressing $a$ and $b$ of Eq.~\eqref{eq:coeff} in terms of the elements and eigenvalues of $\hat{R}$, we derive ${\cal H}$ corresponding to the three categories in the preceding section.

i) When $y_1 \neq y_2$,  setting $y_1 = y$ and $x_1 = \log(y,m) = -x_2$ and then equating the eigenvalues of the matrices in Eq.~\eqref{eq:coeff} simply yield the coefficients $a$ and $b$. Consequently,
\begin{equation} \hat{Z} = \frac{\log(y,m)}{y-1/y} \left( 2 \hat{R} - \frac{y^2+1}{y}\hat{I} \right), \label{eq:z2} \end{equation}
which is indeed traceless on account of $y_1+y_2=(y^2+1)/y=R_1+R_4$. Hence, Eqs.~\eqref{eq:hgeneral} and~\eqref{eq:z2} give the desired result, Eq.~\eqref{eq:h1}.

ii) When $y_1=y_2=y$ and $\hat{R} = y\hat{I}$, setting $x_1 = \log(y,m) = -x_2$ again leads to $x_1 = \log(y,m) = i\pi (4m+1\mp 1)/2 $, in which $\mp$ corresponds to $y=\pm 1$, i.e., $\theta=0$ and $\theta=\pi$ in Eq.~\eqref{eq:log1}, respectively. 
Since $\hat{R}=y\hat{I}$ always produces $a=0$ and $b=y$ in Eq.~\eqref{eq:coeff} for any non-zero $\hat{Z}$,
an arbitrary traceless $2$ by $2$ matrix with its two distinct eigenvalues: $i\pi (4m+1\mp 1)/2 $ and $-i\pi (4m+1\mp 1)/2 $ respectively, is always a valid solution to $\hat{Z}$. Hence, Eq.~\eqref{eq:h2}.

iii) When $y_1=y_2=y$ but $\hat{R}\neq y \hat{I}$, $\hat{R}$ must be similar to a Jordan normal form with the supradiagonal element of $1$, so does $\hat{Z}$. 
Because this non-diagonal Jordan form must imply $x_1=x_2$, the eigenvalues additionally subject to the constraint: $x_1=-x_2$, have to be both zero: $x_1=x_2=0$ and thus $y=e^{x_1} = 1$.
In this case, higher order multiplications in the Taylor series of Eq.~\eqref{eq:ez} all vanish: $\hat{Z}^k = 0$ for any $k=2,3,\cdots$, and only the first two terms survive to give $a=b=1$ in Eq.~\eqref{eq:coeff}. 
Hence, $\hat{Z} = \hat{R} - \hat{I}$ and then Eq.~\eqref{eq:h3}. On the other hand, when $y_1=y_2=-1$ but $\hat{R}\neq - \hat{I}$, no traceless solution to $\hat{Z}$ can be found.

{\it Examples and Discussions.} The transition matrix for the symplectic Euler (E) integrator reads explicitly\cite{Yoshida1993,Donnelly_Rogers2005} 
\begin{equation} \hat{R}_{\rm E}(\tau) =  \begin{bmatrix}   R_1(\tau) & R_2(\tau)  \\ R_3(\tau)  & R_4(\tau) \end{bmatrix}  = \begin{bmatrix}   1-\tau^2 & \tau  \\ -\tau  & 1 \end{bmatrix} . \label{eq:reuler} \end{equation}
According to iii-b), ${\cal H}$ does not exist when $\tau = 2$; otherwise, there are an infinite number of valid Hamiltonians. When $0 < \tau < 2 $,  the explicit expression for ${\cal H}$ follows Eq.~\eqref{eq:h1} and the category i-a)
\begin{equation} {\cal H}(\tau, q, p|m) = \lambda_m  \frac{ p^2 + q^2 - \tau pq }{\tau \sqrt{4-\tau^2} } \label{eq:h11}, \end{equation}
where $\lambda_m = 2m\pi + {\rm acos}( 1 -\tau^2/2)$. Here, it is crucial to use ${\rm acos}( 1 -\tau^2/2) \in (0,\pi) $ rather than ${\rm asin}(\tau\sqrt{ 1 -\tau^2/4}) \in (0,\pi/2)$ for $0 < \tau < 2$.
${\cal H}(\tau, q, p|0)$ at $m=0$ with the smallest $\lvert \lambda_m \rvert$ identifies with the power series of Eq.~\eqref{eq:hseries} and the expression previously derived by Donnelly and Rogers\cite{Donnelly_Rogers2005,coeff_note}.

Similarly, when $\tau > 2$, Eq.~\eqref{eq:h1} and i-c) yield the complex-valued time-independent Hamiltonians:
\begin{equation} {\cal H}(\tau, q, p|m) = \lambda_m  \frac{ p^2 + q^2 - \tau pq }{\tau \sqrt{\tau^2-4} }, \label{eq:h12} \end{equation}
where the complex number $\lambda_m = i(2m+1)\pi+\log 2 - \log(\tau^2 - 2 + \tau\sqrt{\tau^2-4} )$.  
Somewhat surprisingly, the Hamilton's equations of motion in the complex plane do represent the symplectic discrete dynamics in the real space! 
However, the power series of Eq.~\eqref{eq:hseries} yields no valid Hamiltonian any more because it diverges for $\tau > 2$. 
Tracing back to the matrix $\hat{Z}$ in Eq.~\eqref{eq:logR}, it is important to keep its complex-valued solutions; obviously, these solutions can not be derived from the previous BCH technique (e.g.\cite{Yoshida1993}), which at most leads to one real-valued expression.
It should be possible to extend this technique to get multiple power series. Such an extension might be useful for solving other problems that have been dealt with by the BCH technique associated with Lie algebra\cite{Varadarajan1984}.

Setting $t_m = \lambda_m t/\tau$, the trajectory at a given initial condition: $q(0)=q_0$ and $p(0)=p_0$ reads correspondingly,
\begin{equation} \left\{  \begin{aligned} q(t) & =  \left( 2p_0 - \tau q_0 \right)\dfrac{ \sin(t_m) }{\sqrt{4-\tau^2} } + q_0 \cos(t_m) \\ p(t) &=  \left( \tau p_0 - 2q_0 \right)\dfrac{ \sin(t_m) }{\sqrt{4-\tau^2} } + p_0 \cos(t_m)  \end{aligned} \label{eq:ct} \right. \end{equation}
for $0<\tau<2$, and 
\begin{equation} \left\{  \begin{aligned} q(t) & =  \left( 2p_0 - \tau q_0 \right)\dfrac{ \sinh(t_m) }{\sqrt{\tau^2-4} } + q_0 \cosh(t_m) \\ p(t) &=  \left( \tau p_0 - 2q_0 \right)\dfrac{ \sinh(t_m) }{\sqrt{\tau^2-4} } + p_0 \cosh(t_m)  \end{aligned} \right. \label{eq:ct2} \end{equation}
for $\tau > 2$, where the hyperbolic trigonometric functions $\sinh(x)=\left(e^x - e^{-x}\right)/2$ and $\cosh(x)=\left(e^x+e^{-x}\right)/2$.
Eqs.~\eqref{eq:ct} and~\eqref{eq:ct2} have the same structure; the only difference between them is whether the triangle functions or the hyperbolic trigonometric functions are used.
It is easy to verify that both equations agree with the discrete dynamics described by Eqs.~\eqref{eq:dd} and~\eqref{eq:reuler}.
Clearly, the generalized momentum (velocity), $p$, is not the time derivative of the generalized coordinate, $q$, any more: $ p \neq \dot{q}$,
which is evident from the fact that the binomial forms in ${\cal H}(\tau, q, p|m)$ of Eqs.~\eqref{eq:h11} and~\eqref{eq:h12} include the nontrivial coefficients and the cross term of $pq$. 
All of these constitute the correction to the direct output of the total energy, i.e., ${\cal H}_0$.
In addition, it is necessary to derive more accurate expressions of velocities compatible with $q$\cite{Gans_Shalloway2000,Jung2018,Toxvaerd2024}, rather than use the direct output of $p$ from the integrator; and vice versa. 
Other quantities such as kinetic and potential energies are then defined accordingly.
Fig.~\ref{fig:1} shows the first several trajectories obtained from Eq.~\eqref{eq:ct} with $ 0 < \tau = 0.66 < 2$, $m=0, \pm 1$ and the initial state: $(q_0=1,p_0=0)$.
In the $qp$ phase space, the trajectory rotates clockwise for any $m \geqslant 0$ and counter-clockwise for any $m < 0$.  Besides, the period dramatically decreases as $\left| m \right|$ and the real-valued $t_m$ increase when $0<\tau<2$.
In contrast, when $\tau >2$, the hyperbolic trigonometric functions with the complex-valued $t_m$ as the input must result in divergent trajectories.

${\cal H}$ for the usual velocity-Verlet (v) or position-Verlet (p) integrators\cite{Tuckerman1992,Frenkel_Smit2023} identically falls into the categories i-a), iii-b), and i-c) for $0 < \tau < 2$, $\tau = 2$, and $\tau > 2$ respectively. 
Because $R_1 = R_4=1-\tau^2/2$ in the Verlet integrators, Eq.~\eqref{eq:h1} produces the binomial form of ${\cal H}$ without the cross term of $pq$.
However, the nontrivial coefficients different from $1/2$ in ${\cal H}_0(q,p)$ still exist and contribute to the Boltzmann distribution, $\propto e^{-\beta {\cal H}(q,p,\tau|m)}$, 
which should be sampled by symplectic integrators coupled with a thermostat. The coupling possibly follows the current schemes\cite{Leimkuhler_Matthews2013,Liu_Shao2017,therm_note}.

At some critical time-increment, typical symplectic integrators often suffer from the non-existence of ${\cal H}$ due to the non-diagonal Jordan form with eigenvalues $-1$. 
It is always possible to create new symplectic integrators to eliminate this singularity.
For example, one might simply define a double Euler (dE) integrator as $\hat{R}_{\rm dE}(\tau) =  \hat{R}_{\rm E}(\tau/2) \hat{R}_{\rm E}(\tau/2)$ such that its eigenvalues at the critical time-increment, $\tau/2=2$, change from $-1$ to $1$. 
${\cal H}$ for $\hat{R}_{\rm dE}(\tau)$ falls into the categories i-a) for $ 0 <\tau < 2\sqrt{2}$ and $ 2\sqrt{2} <\tau < 4$, ii) for $\tau = 2\sqrt{2}$, iii-a) for $\tau = 4$, and i-b) for $\tau > 4 $, respectively.
However, a combination of two distinct integrators might still fail, just as $\hat{R}_{\rm vp}(\tau) = \hat{R}_{\rm v}(\tau/2)\hat{R}_{\rm p}(\tau/2)$ does. It is of course straightforward to analyze $\hat{R}_{\rm vp}(\tau)$ and any other time-independent transition matrix for the harmonic oscillator. 
Indeed, $\hat{R}_{\rm vp}(\tau)$ improves the accuracy for the direct output of the total energy, because the correction revealed by the corresponding ${\cal H}(\tau, q, p|m)$ of Eq.~\eqref{eq:h1} is order of $\tau^3$ as opposed to ${\cal O}(\tau^2)$ for the traditional velocity- or position-Verlet integrator.

\textcolor{black}{
{\it Shadow Energy of the Verlet integrator}
The propagation of a classical system according to Eq.~\eqref{eq:dd} is a discrete version of the Hamiltonian dynamics defined by Eq.~\eqref{eq:ceh0} --- two coupled first-order differential equations simultaneously govern the time evolution of the pair of variables $(q,p)$ subject to an initial condition: $(q(0),p(0)) = (q_0,p_0)$.
In fact, the widely known Verlet or St{\"o}rmer algorithm\cite{Verlet1967,Levesque_Verlet1993,Delambre1791,Newton1687,verlet_note} that updates the position $x$ according to
\begin{equation} x((n+1)\tau) = 2 x(n\tau) - x((n-1)\tau) + \tau^2 f(n\tau),  \label{eq:vd} \end{equation}
with $f$ the mass-reduced net force, is a discrete version of the Newtonian dynamics defined by $\ddot{x}(t) = f$ --- one second-order differential equation governs the time evolution of the single variable $x$ subject to initial conditions: $x(0)=x_0$ and $\dot{x}(0) = v_0$. 
Parallel to the investigation of the perturbed Hamiltonian\cite{Yoshida1993}, it must be equally fruitful to examine the possibility of an exactly conserved quantity in the Verlet dynamics. 
Indeed, Toxvaerd has coined the term shadow energy and has first discussed its conservation in simple model systems\cite{Toxvaerd1994}.
In recent years, excellent simulations of complex systems\cite{Toxvaerd_Dyre2012,Toxvaerd2012a,Hammonds_Heyes2020,Hammonds_Heyes2021} have followed Toxvaerd's work to calculate the precise value of the shadow energy with an exceptional accuracy of $10^{-10}$ in the reduced unit\cite{Hammonds_Heyes2020}. 
The explicit ${\cal H}(\tau,p,q|m)$ of Eq.~\eqref{eq:h1} in a closed form allows a convenient comparison between the perturbed Hamiltonian and the shadow energy, $E_{\rm s}(\tau,x)$ (e.g. Eq.(4) of ref.\cite{Hammonds_Heyes2020,Hammonds_Heyes2021,typo_note}) for the Verlet dynamics of the harmonic oscillator with $f=-x$.
}

\textcolor{black}{
To this end, the velocity in the Verlet algorithm must be defined since it does not appear explicitly in Eq.~\eqref{eq:vd}. In the so-called leap-frog scheme, the velocity is expressed as the center-difference of the positions (e.g. Eq.(3) of ref.\cite{Hammonds_Heyes2020}):
\begin{equation} v(n\tau) = \left[ x((n+1)\tau) - x((n-1)\tau) \right] /(2\tau) \label{eq:vdv}, \end{equation}
Let $(x,v)=(q,p)$ in the reduced unit, the combination of Eqs.~\eqref{eq:vd} and~\eqref{eq:vdv} identifies with the velocity Verlet integrator\cite{Frenkel_Smit2023} with the elements of the transition matrix: $R_1=R_4=1-\tau^2/2$, $R_2 = \tau$ and $R_3 = -\tau + \tau^3/4$. 
Consequently, for $0 < \tau < 2$, $\lambda_0 = {\rm acos}( 1 -\tau^2/2)$, the same as that for the symplectic Euler integrator and 
\begin{equation} {\cal H}(\tau,q,p|0)= \frac{\lambda_0}{\tau\sqrt{4-\tau^2}} (q^2+p^2-\tau^2 q^2/4). \end{equation}
Therefore, while $q^2+p^2-\tau^2 q^2/4$ is certainly a conserved quantity or conserved `energy', it is not a perturbed Hamiltonian for the discrete dynamics, i.e., the $q$- and $p$-independent coefficient $\lambda_0/(\tau\sqrt{4-\tau^2})$ should not be overlooked whenever the canonical equations of motion are concerned.
}

\textcolor{black}{
Expanding the coefficient to a power series in $\tau$:
\begin{equation} \frac{\lambda_0}{\tau\sqrt{4-\tau^2}} = \sum_{k=0}^\infty \frac{\tau^{2k}\left(k!\right)^2}{2(2k+1)!} = \frac{1}{2}+\frac{\tau^2}{12}+\frac{\tau^4}{60}+\cdots ,\end{equation}
yields the exact series for ${\cal H}(\tau,q,p|0) $:
\begin{multline} {\cal H}(\tau,q,p|0) = {\cal H}_0(q,p) + \frac{\tau^2}{24}\left(2p^2-q^2\right) \\ + \frac{\tau^4}{720}\left(12p^2-3q^2\right) + {\cal O}(\tau^6), \label{eq:hexpand} \end{multline}
compared with the shadow energy $E_{\rm s}$ derived by Hammonds and Heyes\cite{Hammonds_Heyes2020,Hammonds_Heyes2021,typo_note} applied to the harmonic limit,
\begin{multline} E_{\rm s}(\tau,x,v) = {\cal H}_0(x,v) + \frac{\tau^2}{24}\left( -2v \dot{f} - f^2 \right) \\ + \frac{\tau^4}{720} \dot{f}^2 + {\cal O}(\tau^4), \end{multline}
in which the first two terms up to ${\cal O}(\tau^2)$ agree with the expressions in the work by Toxvaerd {\it et al.}\cite{Toxvaerd_Dyre2012,Toxvaerd2012a}
$E_{\rm s}$ with $f=-x=-q$ and $\dot{f}=-p$ differs from ${\cal H}(\tau,q,p|0)$ in the high order term associated with $\tau^4/720$ and this difference is likely responsible for the error of ${\cal O}(\tau^4)$ in $E_{\rm s}$, as opposed to ${\cal O}(\tau^6)$ in Eq.~\eqref{eq:hexpand}, for complex systems as well\cite{Hammonds_Heyes2020}.
}

In a word, we have exactly solved the Hamiltonian representation of arbitrary discrete symplectic dynamics of the one-dimensional single harmonic oscillator.
The representation illustrates the structure, regularity and stability of the discrete symplectic dynamics in a transparent way. 
At a small time-increment, $\tau$, the obtained time-independent Hamiltonian avoids the standard power series derived by means of the Baker-Campbell-Hausdorff (BCH) technique associated with Lie algebra and it is proved to be non-unique.
At a large $\tau$, the complex-valued Hamilton's equations of motion surprisingly represent the discrete motion in the real space. 
\textcolor{black}{
The present solution identically provides the result for a string of independent harmonic oscillators, which commonly models decoupled slight vibrational motions in a solid\cite{Frenkel_Smit2023}.
}
It is also possible to solve any linear system of many coupled harmonic oscillators in the three dimensions, because the derived transition matrice can be simplified as a function of only time-increment. 
For a nonlinear discrete dynamics determined by a transition matrix containing $q$- and $p$-dependent operators,
the convergence of the power series has not been strictly known even for a small time-increment\cite{Yoshida1993}; however, our exact result for the one-dimensional model shows that the series does converge as long as the trajectory is periodic rather than divergent.
This observation implies that for the nonlinear system described by ${\cal H}_0 = q^4 + p^2/2$ or an isolated mechanical system with a finite Poincar\'e recurrence time, the convergence in $\tau$ may be generally valid, but it may not be for an essentially divergent system with unbound nonlinear potentials (e.g. ${\cal H}_0 = q^3 + p^2/2$).
\section*{Acknowledgments}
We are grateful to S{\o}ren Toxvaerd and Haruo Yoshida for their useful communications. One of the authors (Z.H.) appreciates the discussion with Hao Wu and Yong-shi Wu in the Online Club Nanothermodynamica (founded in June 2020). 
\textcolor{black}{
We would also like to acknowledge the anonymous reviewers for their comments to the early versions of the draft, which were helpful to improve the quality of the current work.
}
This work was supported by NSFC (Grant No. 22273047) and no potential conflict of interest was reported by the authors.

\begin{thebibliography}{36}
\bibitem [{coe()}]{coeff_note}%
  \BibitemOpen
  \href@noop {} {}\bibinfo {note} {Note that Eq.(3.23) of
  ref.\cite{Donnelly_Rogers2005} holds only for $0 < \tau \leqslant \sqrt{2} $;
  when $\sqrt{2} < \tau < 2$, the involved ${\rm asin}(\tau\sqrt{1-\tau^2/4})$
  should be replaced by either ${\rm acos}(1-\tau^2/2)$ or $\pi - {\rm
  asin}(\tau\sqrt{1-\tau^2/4})$.}\BibitemShut {Stop}%
\end{thebibliography}

%
\end{document}